\documentclass{article}
\usepackage{amsmath,amsthm}
\usepackage{amssymb,latexsym}
\usepackage[mathscr]{eucal}
\usepackage{setspace}
\usepackage{graphics}
\usepackage{array}

\newcommand{\pa}{\partial}
\newcommand{\va}{\varepsilon}
\newcommand{\om}{\omega}
\newcommand{\Om}{\Omega}

\newcommand{\al}{\alpha}
\newcommand{\na}{\nabla}
\newcommand{\de}{\delta}

\newcommand{\lp}{\left(}
\newcommand{\rp}{\right)}
\newcommand{\lb}{\left[}
\newcommand{\rb}{\right]}

\newcommand{\rc}{\right\}}

\newcommand{\beq}{\begin{equation}}
\newcommand{\eq}{\end{equation}}
\newcommand{\bs}{\boldsymbol}
\newcommand{\ti}{\times}
\newcommand{\fhalf}{\frac{1}{2}}

\begin{document}

\title{\textbf{Plasma Relaxation and Topological Aspects in Electron Magnetohydrodynamics}}         
\author{B. K. Shivamoggi\\
University of Central Florida\\
Orlando, FL 32816-1364, USA\\
}        
\date{}          
\maketitle

\noindent \large{\bf Abstract} \\ \\

Parker's formulation of isotopological plasma relaxation process toward minimum magnetics energy states in magnetohydrodynamics (MHD) is extended to electron MHD (EMHD). The lower bound on magnetic energy in EMHD is determined by both the magnetic field and the electron vorticity field topologies, and is shown to be reduced further in EMHD by an amount proportional to the sum of total electron-flow kinetic energy and total electron-flow enstrophy. The EMHD Beltrami condition becomes equivalent to the "potential vorticity" conservation equation in two-dimensional (2D) hydrodynamics, and the torsion coefficient and turns out to be proportional to "potential vorticity". The winding pattern of the magnetic field lines appears to evolve therefore in the same way as "potential vorticity" lines in 2D hydrodynamics.

\pagebreak

\noindent\large\textbf{1. Introduction}\\

In the magnetohydrodynamics (MHD) model, ions dominate the dynamics while electrons merely serve to shield out rapidly any charge imbalances. On the other hand, in electron MHD (EMHD) with length scales $\rho_e\ll\ell\ll\rho_i,\rho_s, s=i,e$ being the gyro-radius, electrons dominate the dynamics while the demagnetized ions merely serve to provide the neutralizing static background (Kingsep et al. \cite{Kin}. Gordeev et al. \cite{Gor}). The EMHD model restricts itself further to length scales $l\ll d_s$, where $d_s\equiv c/\om_{ps}$ is the skin depth, and frequencies $\om >\om_{ci}$ and $\om_{pi}, \om_{c_s}$ being the cyclotron frequency. The frozen-in condition of magnetic field in EMHD is destroyed by electron inertia. Observations of plasmas in the magnetosphere (Deng and Matsumoto \cite{Den}) and laboratory (Ren et al. \cite{Ren}) showed that magnetic reconnection process is initiated in very thin current sheets (thickness $\sim 0 \lp d_e\rp$).

Equations governing several plasma dynamics models are known to admit a significant class of exact solutions (Shivamoggi \cite{Shi}) under the Beltrami condition - the local current density is proportional to the magnetic field - the force-free state (Lundquist \cite{Lun}, Lust and Schluter \cite{Lu}). Physically, the Beltrami solutions reflect the magnetic topology (in particular, the magnetic flux linkage) constraint on the ways in which plasmas can minimize total energy and are known to correlate well with real plasma behavior (Priest and Forbes \cite{Pri}, Schindler \cite{Sch}). On the other hand, Parker \cite{Par}-\cite{Par3} showed that, in certain plasma processes, the Beltrami condition is indeed equivalent to the vorticity conservation equation in two-dimensional (2D) hydrodynamics (and the Lagrange multiplier $\alpha$ turned out to be proportional to vorticity).
	An important issue associated with the role of current sheets in the development of plasma equilibrium in the EMHD model is the relaxation of plasma as the magnetic field lines are wrapped around and intermixed by turbulent motions in the plasma. (Current-sheet formation may indeed be viewed as a natural concomitant of magnetic relaxation to a minimum energy state (Parker \cite{Par4})). In particular, it is important to determine if the current sheets play the same role in the development of EMHD equilibrium as they do in MHD (Parker  \cite{Par}-\cite{Par3}) and Hall MHD (Shivamoggi \cite{Shi2}), and how the winding pattern of the magnetic field lines evolves in EMHD. The purpose of this paper is to address this issue.

\vspace{.20in}

\noindent\large\textbf{2. Beltrami States in EMHD}\\

On assuming that the displacement current  $\pa{\bf E}/\pa t$ is negligible, which is valid if $\om\ll\om^2_{p_e}/\om_{c_e}$, the continuity of electron flow implies (in usual notation),

\beq\notag
\frac{\pa n_e}{\pa t}=-\na\cdot\lp n_e{\bf v}_e\rp=\na\cdot\left(\frac{{\bf J}}{e}\right)=\frac{c}{e}\na\cdot\lp \na\ti{\bf B}\rp=0
\eq

\noindent which leads to the assumption underlying the EMHD model,

\beq
n_e=const.
\eq
	
\noindent The EMHD equations are summarized by

\beq
\frac{\pa{\bs\Om}e}{\pa t}=\na\ti\lp{\bf v}_e\times{\bs\Om}_e\rp
\eq

\noindent where ${\bs\Om}_e$ is the generalized electron-flow vorticity 

\beq
{\bs\Om}_e\equiv{\bs\om}_e+{\bs\om}_{c_e},{\bs\om}_e\equiv\na\ti{\bf v}_e,{\bs\om}_{c_e}\equiv -\frac{e{\bf B}}{m_e c}.
\eq
	
Equation (1) has the Hamiltonian formulation (Shivamoggi \cite{Shi}),

\beq\notag
H=\fhalf{\underset{V}{\int}}\lp m_e n_e{\bf v}_e^2+{\bf B}^2\rp dV
\eq

\noindent or

\beq\notag
H=\fhalf{\underset{V}{\int}}\lp {\bs\psi}_e\cdot {\bs\om}_e+\frac{1}{c}{\bf A}\cdot{\bf J}\rp dV
\eq

\noindent or

\beq
H=\fhalf{\underset{V}{\int}}{\bs\psi}_e \cdot{\bs\Om}_e dV
\eq

\noindent where,

\beq
m_e n_e {\bf v}_e\equiv\na\ti{\bs\psi}_e
\eq

\noindent which is compatible with (1). We assume either that ${\bf\hat n}\cdot{\bs\Om}_e=0$ on a surface $S$ which bounds the volume $V$ occupied by the plasma and moves with the electron fluid or that $V$ is unbounded and ${\bs\Om}_e$ falls away sufficiently rapidly. The non-uniqueness implicit in (5) may be resolved via the gauge condition,

\beq
\na\cdot{\bs\psi}_e=0.
\eq
	
We take ${\bs\Om}_e$ to be the canonical variable and the skew-symmetric transformation producing differential operator $J$ to be 

\beq
J\equiv-\na\ti\lb\lp\frac{{\bs\Om}_e}{m_en_e}\rp\ti\lp\na\ti\lp\cdot\rp\rp\rb.
\eq

Hamilton`s equation is then

\beq
\frac{\pa{\bs\Om}_e}{\pa t}=J\frac{\de H}{\de{\bs\Om}_e}
\eq

\noindent or

\beq\notag
\begin{matrix}
\begin{aligned}
\frac{\pa{\bs\Om}_e}{\pa t}&=-\na\ti\lb\lp\frac{{\bs\Om}_e}{m_en_e}\rp\ti\lp\na\ti{\bs\psi}_e\rp\rb\\
\\
&=\na\ti\lp{\bf v}_e\ti{\bs\Om}_e\rp
\end{aligned}
\end{matrix}
\eq

\noindent as required. Here, $\de H/\de q$ is the variational derivative.
	
The Casimir invariant for EMHD is a solution of the equation,

\beq
J\frac{\de\mathscr{C}}{\de{\bs\Om}_e}=-\na\ti\lb\lp\frac{{\bs\Om}_e}{m_en_e}\rp\ti\lp\na\ti\frac{\de\mathscr{C}}{\de{\bs\Om}_e}\rp\rb={\bf 0}
\eq

\noindent from which,

\beq
\frac{\de\mathscr{C}}{\de{\bs\Om}_e}={\bf v}_e-\frac{e{\bf A}}{m_e c}
\eq

\noindent so, 

\beq
\mathscr{C}={\underset{V}{\int}}\lp{\bf v}_e-\frac{e{\bf A}}{m_e c}\rp\cdot{\bs\Om}_e~ dV\equiv H_e
\eq

\noindent which is the total generalized electron-flow magnetic helicity $H_e$. The invariance of $\mathscr{C}$ appears to signify some restrictions on the topological aspects of magnetic field and electron vorticity in EMHD (Shivamoggi \cite{Shi}).
	
The Beltrami state in EMHD, which is the minimizer of $H$ keeping $\mathscr{C}$ constant, is given by

\beq
\frac{\de H}{\de{\bs\Om}_e}=\lambda\frac{\de\mathscr{C}}{\de{\bs\Om}_e}
\eq

\noindent or

\beq
{\bs\psi}_e=\lambda\lp{\bf v}_e-\frac{e{\bf A}}{m_e c}\rp
\eq

\noindent or

\beq
m_en_e{\bf v}_e=\lambda{\bs\Om}_e
\eq

\noindent which may be rewritten as

\beq
d_e^2\na\ti\lp\na\ti{\bf B}\rp-\al\na\ti{\bf B}+{\bf B}=0
\eq

\noindent where $d_e$ is the electron skin depth

\beq\notag
d_e\equiv c/\om_{p_e}.
\eq

\vspace{.15in}
\noindent\large\textbf{3. Plasma Relaxation in an Applied Uniform Magnetic Field: Parker Problem in EMHD}\\

Following Parker  \cite{Par}-\cite{Par3}, consider a plasma in an applied magnetic field and confined between two infinite parallel planes $z=0$ and $L$. The field lines of an initially uniform magnetic field ${\bf B}_0=B_0{\bf\hat i}_z$ are wrapped around and intermixed by random turbulent motion of their footpoints on these two planes. This interlaced magnetic field then relaxes under the control of topological aspects toward the lowest available energy state\footnote{As a solar plasma application, magnetic field lines at the edge of the photosphere execute a random walk due to turbulent convection in the photosphere. These braided magnetic field lines are believed to relax isotopologically toward minimum magnetic energy states. The lower bound on magnetic energy is determined by the magnetic field topology, because a non-trivial magnetic topology may be seen to hinder the full dissipation of magnetic energy (Arnol'd and Khesin \cite{Arn}). This lower bound is reduced further in EMHD by amount proportional to the sum of the total electron-flow kinetic energy and total electron-flow enstrophy (see Appendix).} described by

\beq
d_e^2\na\ti\lp\na\ti{\bf B}\rp-\al\na\ti{\bf B}+{\bf B}=0
\eq

\noindent where the Lagrange multiplier $\al$ may be interpreted as a torsion coefficient.

Suppose this process exhibits slow variations in the $z$-direction, characterized by the slow spatial scale

\beq
\xi\equiv \va z, \phantom{x}\va\ll 1
\eq

\noindent Let the magnetic field involved in this process be given by

\beq
{\bf B}=\va B_0b_x{\bf\hat i}_x+\va B_0b_y{\bf\hat i}_y+B_0\lp 1+\va bz\rp{\bf\hat i}_z
\eq

\noindent and let the Lagrange multiplier $\al$ be given by

\beq
\al=1/\va a.
\eq

Using (16)-(18), equation (14) may be rewritten as

\beq\tag{19a}
v_{ex}=\sigma\lp C_1\va b_x+\om_{ex}\rp\approx\sigma\lp C_1\va b_x+\frac{\pa v_{ez}}{\pa y}\rp
\eq

\beq\tag{19b}
v_{ey}=\sigma\lp C_1\va b_y+\om_{e_y}\rp\approx\sigma\lp C_1\va b_y+\frac{\pa v_{ez}}{\pa x}\rp
\eq

\beq\tag{19c}
v_{ez}=\sigma\lb C_1\lp 1+\va b_z\rp+\om_{e_z}\rb
\eq

\noindent where $\sigma$ and $C_1$ are appropriate constants.

On the other hand, equation (15) may be rewritten as

\beq\tag{20a}
\frac{\pa b_z}{\pa y}-\va\frac{\pa b_y}{\pa \xi}=\va a\lp b_x-d_e^2\na^2b_x\rp
\eq

\beq\tag{20b}
\va\frac{\pa b_x}{\pa \xi}-\frac{\pa b_z}{\pa x}=\va a\lp b_y-d_e^2\na^2b_y\rp
\eq

\beq\tag{20c}
\frac{\pa b_y}{\pa x}-\frac{\pa b_x}{\pa y}= a\lb\lp 1+\va b_z\rp-\va d_e^2\na^2b_z\rb.
\eq

\noindent The divergence - free condition on ${\bf B}$ gives

\beq\tag{21}
\frac{\pa b_x}{\pa x}+\frac{\pa b_y}{\pa y}+\va\frac{\pa b_z}{\pa\xi}=0.
\eq

\noindent Equations (20a) and (20b) imply, 

\beq\tag{22}
b_z\sim 0\lp\va\rp
\eq

\noindent Using (22), equation (21) becomes

\beq\tag{23}
\frac{\pa b_x}{\pa x}+\frac{\pa b_y}{\pa y}\approx 0.
\eq

\newpage
\noindent which implies,

\beq\tag{24}
b_x=\frac{\pa \psi}{\pa y},~ b_y=-\frac{\pa\psi}{\pa x}
\eq

\noindent for some $\psi=\psi\lp x,y\rp.$

Using (24), equation (20c) leads to 

\beq\tag{25}
a=-\na^2\psi
\eq

\noindent while equations (19a) and (19b) lead to 

\beq\tag{26a}
\frac{\pa v_{e_x}}{\pa y}\approx\sigma\left(C_1\va\frac{\pa^2 \psi}{\pa y^2}+\frac{\pa^2 v_{e_z}}{\pa y^2}\right)
\eq

\beq\tag{26b}
\frac{\pa v_{e_y}}{\pa x}\approx\sigma\left(-C_1\va\frac{\pa^2 \psi}{\pa x^2}-\frac{\pa^2 v_{e_z}}{\pa x^2}\right).
\eq

Using equations (26a) and (26b), we obtain

\beq\tag{27}
\omega_{ez}=\frac{\pa v_{e_y}}{\pa x}-\frac{\pa v_{e_x}}{\pa y}\approx-\va\sigma\left(C_1\na^2\psi+\frac{1}{\va}\na^2 v_{e_z}\right).
\eq

Put, 

\beq\tag{28}
\left.
\begin{matrix}
\begin{aligned}
v_{e_z}&\equiv C_1\lp \sigma+\va w\rp\\
\\
\omega_{e_z}&\equiv\va\sigma C_1\omega
\end{aligned}
\end{matrix}
\rc
\eq

\noindent we then obtain from equation (19c),

\beq\tag{29}
w\approx\sigma^2\om
\eq

Using (28) and (29), equation (27) gives

\beq\tag{30}
\om\approx-\na^2\lp\psi+\sigma^2\om\rp.
\eq

Using (30), equation (25) leads to

\beq\tag{31}
a=\om-\sigma^2\na^2\om\sim q.
\eq

\noindent So, the torsion coefficient $a$ is proportional to the ``potential vorticity'' $q$.

On the other hand, taking the divergence of equation (15), we obtain

\beq\tag{32}
\lp\na\ti{\bf B}\rp\cdot\na\al=0
\eq

Using (18) and (31), equation (32) leads to 

\beq\tag{33}
v_{ex}\frac{\pa q}{\pa x}+v_{ey}\frac{\pa q}{\pa y}+\va v_{ez}\frac{\pa q}{\pa \xi}=0
\eq

So, the Beltrami condition (15) in EMHD becomes equivalent to the ``potential vorticity'' conservation equation\footnote{By contrast, in Hall MHD (Shivamoggi \cite{Shi2}), the ``potential vorticity'' $q$ therefore becomes identical (apart from a proportionality constant) to the potential vorticity for fluids in the quasi-geostrophic approximation. (Quasi-geostrophic dynamics refers to the nonlinear dynamics governed by the first-order departure from the linear geostrophic balance between the Coriolis force and pressure gradient transverse to the rotation axis of a rapidly rotating fluid.) This may be seen by noting that, using (23c) and (33) in \cite{Shi2}, (34) may be rewritten as

\beq\tag{I}
q\approx\om-\sigma^2\na^2\om
\eq

\noindent The second term on the right in (I) represents an additional transport mechanism for the magnetic field via the Hall current in Hall MHD (Sonnerup \cite{Son}).

Noting from (24c) in \cite{Shi2} that

\beq\tag{II}
\om\sim b_z\eq

\noindent (I) becomes

\beq\tag{III}
q\sim\na^2 b_z-\frac{1}{\sigma^2}b_z
\eq

For comparison, one may note one version of potential vorticity for fluids in quasic-geostrophic approximation (Charney \cite{Cha}),

\beq\tag{IV}
q=\na^2\psi-\frac{1}{l^2}\psi
\eq

\noindent $\psi$ being the stream function of the flow, and $l$ is the Rossby deformation radius. The second term on the right in (IV) represents the  vortex stretching effect due to the deformed free surface in fluid subject to gravity. The similarity between (III) and (IV) becomes all the more striking, especially on noting that the out-of-plane magnetic field $b_z$ plays the role of the stream function for in-plane ion flows in Hall MHD.} in 2D hydrodynamics, and the torsion coefficient. $\al$ turns out to be proportional, as per (31), to ``potential vorticity''. The winding pattern of the magnetic field lines in EMHD therefore evolves in the same way as ``potential vorticity'' lines in 2D hydrodynamics.
 
\vspace{.20in}

\noindent\large\textbf{4. Discussion}\\
	
In this paper, we have extended Parker's  \cite{Par}-\cite{Par3} formulation of isotopological plasma relaxation process toward minimum magnetic energy states in MHD to EMHD. The lower bound on magnetic energy in EMHD is determined by both the magnetic field and the electron-flow vorticity field topologies. This lower bound is reduced further in EMHD by an amount proportional to the sum of total electron-flow kinetic energy and total electron-flow enstrophy (see Appendix). The EMHD Beltrami condition becomes equivalent to the ``potential vorticity'' conservation equation in 2D hydrodynamics, and the torsion coefficient. $\al$ turns out to be proportional to ``potential vorticity".  The winding pattern of the magnetic field lines therefore evolves in EMHD in the same way as ``potential vorticity'' lines in 2D hydrodynamics. The analogy between a smooth, continuous magnetic field in EMHD and the potential vorticity field in 2D hydrodynamics as that in ordinary MHD (Parker \cite{Par3}) implies that the current sheets seem to have the same role in the development of EMHD equilibria as they do in the MHD (and Hall MHD \cite{Arn}) cases. The current sheets are caused by the interaction between the magnetic pressure and magnetic tension. The spontaneous formation of current sheets as a consequence of relaxation to equilibria in EMHD, as in MHD, may be traced to the common form of Maxwell's magnetic stress tensor in both systems (Eugene Parker, private communication). 

\vspace{.20in}
\newpage
\noindent\large\textbf{Appendix}\\

The conservation of generalized magnetic field topology characterized by $H_e$ (see (11)) while minimizing $H$ (see (4)) prevents the latter from decaying to zero in an EMHD relaxation process. This may be seen by following along the lines of Arnol'd \cite{Arn2}.

The total generalized electron-flow magnetic helicity invariant $H_e$ may be rewritten as,                           

\beq\tag{A.1}                                                     
H_e=\int_V{\bf A}_e\cdot{\bf B}_e dV=const
\eq

\noindent where $B_e$ is the generalized magnetic field, 

\beq\tag{A.2}
B_e\equiv\na\ti{\bf A}_e,~{\bf A}_e\equiv {\bf A}-\frac{m_e c}{e} {\bf v}_e.                                         
\eq

Consider minimum energy states that can be attained under the evolution governed by equation (2). We have by Schwarz's inequality, 

\beq\tag{A.3}
\int_V{\bf A}^2_e  dV\cdot\int_V{\bf B}^2_e  dV\leq\left(\int_V {\bf A}_e \cdot {\bf B}_e dV\right)^2 = H_e^2	
\eq
Using Poincare's inequality, 

\beq\tag{A.4}
\frac{\int_V{\bf B}_e^2  dV}{\int_V{\bf A}_e^2  dV}\geq k^2	
\eq

\noindent which insures a positive minimum for generalized magnetic energy, and noting (A.1) and (A.3), we observe that this minimum further connects with topological accessibility,

\beq\tag{A.5} 
H=\fhalf \int_V{\bf B}_e^2  dV\geq k|H_e |.	
\eq

\noindent(A.5) indicates that the topographical barrier provided by the linkage of generalized magnetic field lines, implied by $H_e\not=0$, gives rise to a lower bound on the generalized magnetic energy\footnote{Freedman \cite{Fre} pointed out that, for a closed Riemannian 3-manifold, any non-trivial linking of a divergence-free vector field gives rise to a lower-limit on the energy.} and prevents it from decaying to zero during an EMHD relaxation process. (A.5) leads further to, 
	
$$\int_V{\bf B}^2  dV\geq k|H_e |-2d_e^2 \int_V\lp\na\ti{\bf B}\rp^2  dV-d_e^4 \int_V\lp\na^2 {\bf B}\rp^2 dV$$
	
\noindent or

$$\int_V{\bf B}^2  dV \geq k|H_e|-2 m_e n_e \int_V {\bf v}_e^2 dV-2d_e^2 m_e n_e \int_V {\bs\om}_e^2  dV$$

\noindent or

\beq\tag{A.6}
E_M \geq k|H_e |-2 E_K-2d_e^2 m_e n_e W                                          
\eq

\noindent where $E_M$ is the total magnetic energy, $ E_K$ is the total electron flow kinetic energy, and $W$ is the total electron-flow enstrophy. (A.6) implies a reduction of the lower bound on the magnetic energy in EMHD by an amount proportional to the sum of total electron-flow kinetic energy and total electron-flow enstrophy.

\vspace{.20in}

\noindent\large\textbf{Acknowledgements}\\

\noindent This work was carried out during my stay at the Institute for Fusion Studies, University of Texas, Austin. I am thankful to Professor Swadesh Mahajan for the hospitality as well as discussions. I am thankful to Professor Eugene Parker for his helpful suggestions.

\end{document}